\documentstyle[prb,twocolumn,aps]{revtex}
\input{psfig}
\preprint{}
\begin{document}
\draft
\wideabs{
\title{Time-reversal symmetry breaking at Josephson tunnel junctions
  of purely $d$-wave superconductors}
\author{T. L\"ofwander, V. S. Shumeiko, and G. Wendin}
\address{Department of Microelectronics and Nanoscience,
School of Physics and Engineering Physics,\\
Chalmers University of Technology and G\"oteborg University,\\
S-412 96 G\"oteborg,
Sweden}
\date{\today}
\maketitle


\begin{abstract}
  
  We study spontaneous time-reversal symmetry breaking at Josephson
  tunnel junctions of $d$-wave superconductors in the absence of
  subdominant components of the order parameter. For tunnel junctions,
  when the orientation is close to $0/45$ (for which a gap lobe points
  towards the junction on one side and a gap node on the other), the
  mechanism of the symmetry breaking is the splitting of midgap states
  (MGS) by spontaneous establishment of a phase difference $\phi=\pm
  \pi/2$ across the junction. This occurs for transparencies
  $D\gg\xi_0/\lambda$ and temperatures $k_BT\ll D\Delta_0$, where
  $\xi_0$ is the coherence length, $\lambda$ is the penetration depth,
  and $\Delta_0$ is the maximum energy gap. On the other hand, tunnel
  junctions with $D\ll\xi_0/\lambda$ effectively behave as surfaces,
  for which the mechanism of symmetry breaking is self-induced Doppler
  shifts of MGS. For this instability, we calculate the phase
  transition temperature $k_BT_{TRSB}=(1/6)(\xi_0/\lambda)\Delta_0$
  and show that the spatial shape of the gap is unimportant.

\end{abstract}

\pacs{PACS numbers: 74.50.+r, 74.20.-z}
}
\narrowtext

The possibility of spontaneous time-reversal symmetry breaking (TRSB)
states at various surfaces and interfaces of $d$-wave superconductors
has been intensively studied during the last few
years.\cite{Sigreview} Some experimental findings, such as fractional
flux quanta at certain grain boundaries\cite{Kirtley,Mannhart} and
splitting of the zero-bias conductance peak in zero magnetic
field,\cite{cov,Krupke} have been interpreted as realizations of this
state. Theoretically, several different systems have been under
consideration: surfaces,\cite{MatShi,FRS} twin boundaries,\cite{twin}
and Josephson
junctions\cite{YipPRB95,SigGB,HvOS,Zago,YipLTP97,FYK,FY,Ost} with
special orientations.

\begin{figure}[p]
\centerline{\psfig{figure=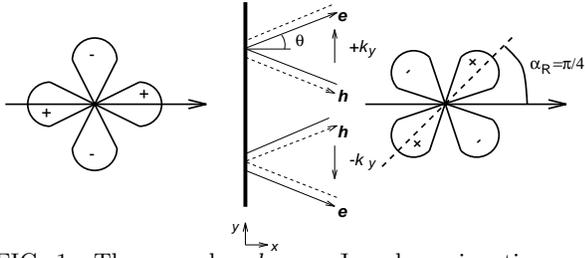,width=7.7cm}}
\caption{The specular $d$-wave Josephson junction under
  consideration: the right superconductor has a gap node pointing
  towards the junction ($\alpha_R=\pi/4$), while the left
  superconductor has a gap lobe pointing towards the junction
  ($\alpha_L=0$). For the surface problem we have vacuum for $x<0$.
  Shown are also the scattering events, consecutive normal scattering
  at the junction and Andreev reflection in the superconductor,
  leading to the formation of midgap states.}\label{fig1}
\end{figure}

In this paper we will study TRSB at Josephson tunnel junctions with
orientation close to $\alpha_L=0$, $\alpha_R=\pi/4$ (Fig.~\ref{fig1})
for arbitrary transparency $D$ of the tunnel barrier. We emphasize
that we are considering purely $d$-wave superconductors, meaning that
a sub-dominant component of the order parameter is assumed to be {\em
  absent}. As will be shown, the TRSB effect is due to the specific
properties of the midgap states formed in these structures.

In 1994 Hu\cite{Hu} showed that surface states with zero energy,
so-called midgap states (MGS), are formed at surfaces and interfaces
of $d$-wave superconductors if the orientation angle $\alpha$ is
non-zero. The largest spectral weight of the MGS appears when a
$d$-wave gap node points directly towards the surface/interface
($\alpha=\pi/4$).

It has been pointed out (see e.g. Refs.~\onlinecite{Sigreview,FY})
that the large density of states exactly at the Fermi level associated
with the MGS is energetically unfavorable: if there exist mechanisms
able to shift the MGS and produce a gap in the spectrum, the energy
will be lowered and a phase transition into a state with broken
time-reversal symmetry will take place. Splitting of MGS due to a
complex $d+is$ order parameter, with a subdominant surface/interface
$s$-wave component, has been considered both at free
surfaces\cite{MatShi,FRS} and at Josephson junctions.\cite{FYK,FY} The
possibility of instabilities at Josephson junctions of purely $d$-wave
superconductors was first pointed out by Yip\cite{YipPRB95} for the
weak link case (transparency $D=1$, no backscattering in the junction,
see also Refs.~\onlinecite{HvOS,Zago}). The TRSB state was later shown
to be favorable also for finite but rather high transmissivity of the
junction,\cite{YipLTP97,FYK,FY} $0.3 \alt D \leq 1$. According to the
symmetry arguments presented in
Refs.~\onlinecite{YipPRB95,YipLTP97,FYK,FY}, the effect of TRSB in
pure $d$-wave junctions heavily relies upon the non-sinusoidal
current-phase relation in transparent weak links. Since the higher
harmonics in the current-phase relation disappear in the tunnel limit,
TRSB in tunnel junctions of pure $d$-wave superconductors is also
expected to disappear and it was supposed that TRSB can only occur
under such circumstances when a complex order parameter is
formed.\cite{FYK,FY} However, it turns out that the symmetry argument
holds only for continuum states and for finite energy bound states,
but not for the MGS which contribute to the Josephson current with a
term proportional to the first power in $D$ at low
temperature.\cite{Tanaka_dcJE,Bagwell_PRB98} In the low-transparency
limit it is not the non-sinusoidal current-phase relation itself that
is important, but rather the uneven occupation of split MGS. Here we
will show that this leads to TRSB also in tunnel junctions of purely
$d$-wave superconductors. The driving mechanism of the instability is
the displacement of MGS induced by spontaneous establishment of a
finite phase difference $\phi=\pm\pi/2$ across the junction, similar
to what happens in transparent
junctions.\cite{YipPRB95,YipLTP97,FYK,FY} In the extreme tunnel limit,
the mechanism crosses over to self-sustained Doppler shifts of MGS,
which also produces instabilities at free
surfaces.\cite{Hig,HonSig,BKK_prb00}

Although the MGS contribution does not appear in conventional tunnel
model calculations (despite the fact that it is proportional to the
first power in transparency $D$), the quasiclassical Green's function
technique, in principle, includes it.  However, since the necessary
condition for TRSB for low transparency is $k_BT\ll D\Delta_0$
(corresponding to uneven occupation of split MGS, see below), the
specific window of transparencies favoring TRSB found in
Refs.~\onlinecite{YipLTP97,FY}, $0.3 \alt D \leq 1$, was actually a
result of the choice of temperature, $k_BT=0.2k_BT_c\sim0.1\Delta_0$.

Consider now the Josephson tunnel junction in Fig.~\ref{fig1}. We
model the junction between the two clean two-dimensional $d$-wave
superconductors by a square specular barrier. In this case, the
quasiparticle wavefunctions can be labeled by the conserved wave
vector component parallel to the surface, $k_y=k_F\sin\theta$, where
$k_F$ is the Fermi wave vector. The gap functions in the
superconductors are
$\Delta_{L/R}(x,\theta) = \Delta_0 g_{L/R}(x)
\cos[2(\theta-\alpha_{L/R})]$,
where all angles are measured relative to the surface normal
(Fig.~\ref{fig1}). The fact that the gap may be suppressed near
surfaces and interfaces of $d$-wave superconductors is reflected in
the $x$-dependent functions $g_{L/R}(x)$.

\begin{figure}[p]
\centerline{\psfig{figure=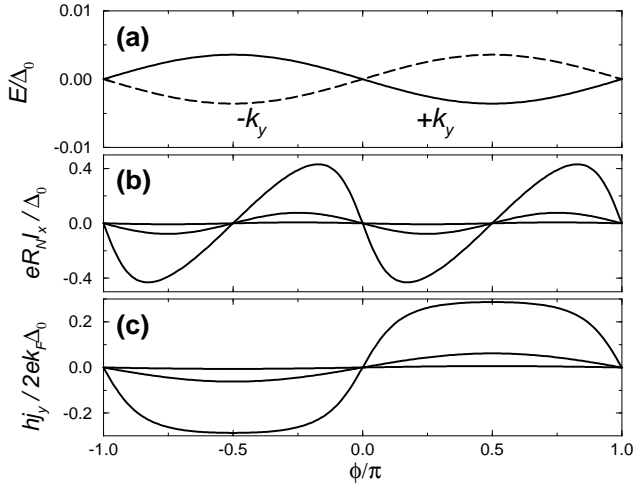,width=8cm}}
\caption{(a) Dispersion of the Andreev bound state with the phase
  difference across the junction. Solid and dashed lines are bound
  states at angles $\theta=\pi/9$ and $\theta=-\pi/9$ respectively,
  $D(\theta=\pm \pi/9)\approx 0.01$. (b) The current-phase relation at
  three different temperatures: $T=0.1T_c$, $T=0.01T_c$, and
  $T=0.001T_c$, for $D\approx 0.009$ ($k_Fd=5$, $U=1.2E_F$, where $U$
  and $d$ are the height and width of the barrier and $E_F$ is the
  Fermi energy). For decreasing temperature, the current crosses over
  from being of order $D^2$ to $D$ and becomes increasingly
  non-sinusoidal. (c) Phase dependence of the surface current density
  calculated to the right of the barrier.}\label{fig2}
\end{figure}

For the calculation of the dc Josephson current, one needs to consider
contributions both from continuum states and Andreev bound states. If
the two superconductors were decoupled (zero transparency), there
would be midgap surface states on the right side ($\alpha_R=\pi/4$)
and also finite energy surface states since the gap is suppressed near
the surface forming a quantum well. In addition there are gap edge
states on the left side ($\alpha_L=0$). For finite transparency the
surface states form states of the entire junction. The energy of these
Andreev states are shifted relative to the surface levels, the shift
depending on the transparency of the barrier and the phase difference
$\phi$ across the junction. To clearly see the mechanism of the TRSB
instability we first consider a step-function dependence of the gap,
$g_{L/R}(x)=\Theta(\mp x)$. By solving the quasiclassical
Bogoliubov-de Gennes equation for the junction, we find the energy of
the midgap state (the $\theta$-dependence of the order parameter is
not explicitly written out here)
\begin{eqnarray}
E(\phi,k_y) &=&
\frac{-\mbox{sgn}(k_y)\Delta_L|\Delta_R| D(\theta) \sin\phi}
{2|\Delta_L|
+D(\theta)\left[|\Delta_R|-|\Delta_L|\right]}
+O(D^3)\nonumber\\
&=&-\mbox{sgn}(k_y) E_0(\theta) \sin\phi+O(D^3),
\label{bs}
\end{eqnarray}
as plotted in Fig.~\ref{fig2}(a). When a phase difference is applied
across the junction, the degeneracy of the $\pm k_y$ MGS is lifted, as
emphasized by the solid and dashed lines in Fig.~\ref{fig2}(a). The
contribution to the dc Josephson current from Andreev states is found
via the relation $I_x=(2e/\hbar)(dE/d\phi)n_F(E)$, where $n_F$ is the
Fermi distribution function. The current carried by the bound states
in Eq.~(\ref{bs}) is therefore
\begin{eqnarray}
eR_N I_x^{MGS}(\phi) &=& -\frac{2\pi}{D}
\int_{0}^{1} d\eta E_0(\eta) \cos\phi\nonumber\\
&&\times\tanh\left[\frac{E_0(\eta) \sin\phi}{2k_BT}\right]
+O(D^2),
\label{Ix}
\end{eqnarray}
where $T$ is the temperature, $R_N=\pi h/e^2k_FL_yD$ is the normal
state resistance of the junction, $L_y$ is the junction width,
$\eta=\sin\theta$, and $D=\int d\eta D(\eta)/2$. From Eq.~(\ref{Ix})
we find that the MGS contribution to the Josephson current at low
temperature, $k_BT\ll |E_0|\sim D\Delta_0$, is of {\em first order in
  the transparency $D$}. In what concerns the continuum, $\pm k_y$
states are degenerate and carry current in opposite directions, which
results in a cancellation of the main (of order $D$) current.  The
residual continuum contribution is small, of order $D^2$. This
cancellation is due to the sign change of $\Delta_R$ when we let
$k_y\rightarrow -k_y$, see Fig.~\ref{fig1}.  Both $k_y$ and $-k_y$
contributions have $\sin\phi$ dependences, but they are $\pi$-shifted
relative to each other (the sign of $\Delta_R$ is equivalent to a
phase $\pi$) and carry current in opposite directions. This is the
symmetry argument\cite{YipPRB95,FYK,FY} referred to above. Also the
$\pm k_y$ bound states near the gap edges carry current in opposite
directions.  Although they are not degenerate (like the MGS they split
under phase bias) they are equally populated for $k_BT\ll\Delta_0$ and
the sum of $\pm k_y$ currents is of order $D^2$. We would like to
emphasize that the MGS do not obey the $\pm k_y$ symmetry at low
temperature because MGS with opposite signs of $k_y$ disperse on
opposite sides of the Fermi level and are unequally populated. In
Fig.\ref{fig2}(b) we plot the total current including the dominant MGS
contribution and the small contributions from the gap edge states and
the continuum states.  In agreement with previous work,
\cite{YipPRB95,SigGB,HvOS,Zago,FYK,FY,Ost,YipLTP97,Tanaka_dcJE,Bagwell_PRB98}
the current-phase relation is $\pi$-periodic. For increasing
temperature, the MGS contribution is reduced, and when $k_BT\gg
D\Delta_0$ [the $T=0.1T_c$ curve in Fig.~\ref{fig2}(b)] the current is
small, of order $D^2$. In the intermediate region, the current in
Eq.~(\ref{Ix}) scales as
$D^2/T$.\cite{Tanaka_dcJE,Bagwell_PRB98,Barash_PRL96} The above
arguments leading to the dominating, of order $D$, MGS currents and
small, of order $D^2$, non-MGS currents hold also for general spatial
dependences of the gap functions. However, as shown in
Ref.~\onlinecite{Barash_PRB00}, the numerical prefactor of the current
calculated for step-function gaps is overestimated by a factor about
two, due to an overestimation of the shift of the MGS with phase
difference. However, this will not influence the instability we
discuss in the following.

By a phenomenological argument Yip\cite{YipPRB95} showed (see also the
paper by \"Ostlund\cite{Ost}) that when the $\pm k_y$ symmetry cancels
the first harmonic of the current in purely $d$-wave junctions with
orientation $\alpha_L=0$, $\alpha_R=\pi/4$, time-reversal symmetry is
broken if the parameters of the theory are chosen in such a way that
the coefficient in front of the second harmonic is negative. The
current due to MGS has indeed this negative sign. As a consequence,
the equilibrium phase difference across the junction is
$\phi_{eq}=\pm\pi/2$, since the Josephson energy minimum appears where
the current through the junction is zero and the slope of the
current-phase relation is positive. That these phase differences
really correspond to Josephson energy minima can be understood by
noting that the energy of the midgap state is the lowest for
$\phi_{eq}=\pm\pi/2$, see Fig.~\ref{fig2}(a) and also
Eq.~(\ref{free_energy}). Considering low temperature, when only the
negative energy states are occupied, we see that the degenerate
$\phi_{eq}=\pm\pi/2$ junction states correspond to occupation of $\pm
k_y$ time-reversed MGS, see the illustration in Fig.~\ref{fig1}.
Assuming that the system chooses one minimum ($\phi_{eq}=\pi/2$) or
the other ($-\pi/2$), surface currents in the positive or negative
$y$-direction will appear and time-reversal symmetry is broken. The
spontaneous surface current, calculated via
\begin{equation}\label{jy}
j_y(x) = \frac{e\hbar}{m}
\sum_{{\bf k}} k_y 
\hat\Psi^{\dagger}_{\bf k}(x)\hat\Psi_{\bf k}(x)
n_F(E_{\bf k}),
\end{equation}
where $\hat\Psi_{\bf k}$ are the wavefunctions satisfying the quasiclassical
Bogoliubov-deGennes equation, and $n_F$ is the Fermi distribution
function, is dominated by the MGS for the same reasons as the MGS
dominate the Josephson current. In Fig.~\ref{fig2}(c) we plot
$j_y(\phi)$ calculated to the right of the barrier at $x=0$. The
surface current approaches its maximum value at the equilibrium phase
differences $\phi_{eq}=\pm\pi/2$, but it has opposite signs since MGS
with opposite signs of $k_y$ are occupied at these two phase
differences.

Since the MGS surface current produces a magnetic field which costs
energy, we must include this effect into the discussion of the
instability. The spatial dependence of the magnetic field ${\bf h}$ is
determined by the counterflowing screening currents, and can be
calculated via the superfluid momentum ${\bf p}_s$,
${\bf h}=-(c/e)\nabla\times{\bf p}_s$.
For type II superconductors, like the high-$T_c$ superconductors,
${\bf p}_s$ is found via the London equation
$\nabla^2{\bf p}_s-{\bf p}_s/\lambda^2=
(4\pi e/c^2)
j_y^{MGS}(x)\hat y\equiv f(x)\hat y$.
Because the MGS surface current remains finite in the limit $D=0$, we
are allowed to let $D\rightarrow 0$ and perform the calculation of the
source current with the free surface MGS wavefunction. This allows us
to explicitly take into account the spatial dependence of the gap. The
surface current then takes the form
\begin{equation}\label{jyMGS}
j_y(x)=\frac{e\hbar}{m}\frac{k_F^2}{2\pi}
\int_0^{1} d\eta \,\eta
\frac{e^{-2\zeta}}{N}
\tanh\left(\frac{-E_0(\eta)}{2k_BT}\right),
\end{equation}
where
$\zeta=\int dl |\Delta(l)|/\hbar v_F$,
$x=l\sqrt{1-\eta^2}$,
and
$N=\int_0^{\infty} dx e^{-2\zeta}$
is the normalization constant of the MGS wavefunction. The integration
over trajectory angles $\eta$ is effectively cut off at $\eta<1$
because of the tunneling cone described by $D(\eta)$. Since the MGS
source only has a $y$-component, the non-trivial component of ${\bf
  p}_s$ is the $y$-component which only depends on $x$, ${\bf
  p}_s=p_s(x)\hat y$.  The solution of the differential equation,
satisfying the boundary condition
$\lim_{x\rightarrow\infty}h_z(x)=0$
is
$p_s(x) = b_0 e^{-x/\lambda} +
\lambda \int^x dx' f(x') \sinh \left[(x-x')/\lambda\right]$.
The constant $b_0$ is fixed by the boundary condition
$h_z(x=0)=0$
at the junction,
$b_0 = -\lambda \int_0^{\infty} dx f(x)
\cosh(x/\lambda)$.

An important aspect of the screening problem is the separation of
length scales: the surface current due to the MGS flows within a thin
layer of width $\xi_0$ near the surface, while the screening currents
flow in a much thicker layer of width $\lambda$. For high-$T_c$
superconductors the ratio $\xi_0/\lambda\ll 1$, and all quantities may
be expanded in this small parameter. For this reason the convergence
of the integrals in the expression for $p_s$ is governed by the
function $f(x')$ which decays on the $\xi_0$ length scale. Expanding
the hyperbolic functions, the spatial shape of the gap function is
cancelled in the leading term of the expression for $p_s$: it appears
both in the normalization $N$ of the MGS wavefunction and in the
integrals over $f(x')$ which are integrals over the MGS wavefunction.
Thus, the detailed spatial shape of the gap drops out of the
calculation and the final form of $p_s$ is
\begin{equation}\label{ps0}
p_s(x) = -\frac{\hbar}{\lambda} e^{-x/\lambda}
\int_0^{\eta_1} d\eta \,\eta
\tanh\left(\frac{-E_0(\eta)}{2k_B T}\right)
\left[1+O\left(\frac{\xi_0}{\lambda}\right)\right].
\end{equation}
We are now able to quantitatively study the difference in the
thermodynamic potential $\Omega$ of junctions with and without broken
symmetry: it consists of two parts, the energy cost of having a
magnetic field and the energy gain due to the shifts of MGS
\begin{eqnarray}\label{free_energy}
\Delta\Omega &=&
\int_0^{\infty} dx \frac{h_z^2(x)}{8\pi}
-k_BT\frac{k_F}{\pi}
\int_0^{\eta_1} d\eta \ln \left[\cosh\frac{E_0(\eta)}{2k_BT}\right]\nonumber\\
&=&\frac{k_F\Delta_0}{4\pi}\left[
\frac{\xi_0}{4\lambda}
-\int_0^{1} d\eta \,\eta\sqrt{1-\eta^2}D(\eta)\right],
\end{eqnarray}
where the second line is valid in the low-temperature limit $k_BT\ll
|E_0|\sim D\Delta_0$.  Clearly, for $D\gg\xi_0/\lambda$,
$\Delta\Omega<0$ and there is an instability.

Rotating the superconductors away from the $\alpha_L=0$,
$\alpha_R=\pi/4$ orientation, the equilibrium phase difference across
the junction is shifted continuously away from $\pm\pi/2$ towards
either $0$ or $\pm\pi$ depending on the direction of rotation. This
happens because the $\pm k_y$ symmetry is lost and non-MGS
contributions are able to dominate. However, a numerical calculation
shows that in the low-temperature region, $k_BT\ll D\Delta_0$, where
the MGS dominate, the TRSB, that is $\phi_{eq}\neq 0$ or $\pm\pi$, is
quite robust: it survives rotations up to $10^o$.

From Eq.~(\ref{free_energy}) it is found that when
$D\alt\xi_0/\lambda$ the energy of the magnetic field may become
larger than the Josephson energy and the instability is lost. However,
for such small transparencies it is necessary to take into account the
Doppler shift of the MGS due to the finite superfluid momentum, which
will assist and uphold the TRSB instability. In the low transparency
limit $D\ll\xi_0/\lambda$, for which the Doppler shift is much larger
than the shift due to finite phase difference, the junction behaves
like a free surface with the MGS localized on the right side of the
junction. Using the above calculation but taking the energy of the MGS
to be $E_0(\eta)={\bf p}_s\cdot{\bf v}_F=p_sv_F \eta$, one finds from
Eq.~(\ref{free_energy}) in the zero-temperature limit
$\Delta\Omega=(E_F/16\lambda)(1-4)[1+O(\xi_0/\lambda)]<0$ showing that
the TRSB state is favorable. Eq.~(\ref{ps0}) with $E_0(\eta)=p_sv_F
\eta$ takes the form of a self-consistency equation for $p_s(T)$. Near
the phase transition temperature, $p_s$ is small and the inequality
$p_s v_F<<2k_B T$ is fulfilled; this allows an expansion of the
hyperbolic function, which leads to
$|p_s| = 2\sqrt{5/3}(\hbar/\lambda)\sqrt{1-T/T_{TRSB}}$,
where $k_BT_{TRSB}=(1/6)(\xi_0/\lambda)\Delta_0$ is the temperature
where the second order phase transition into the TRSB state occurs.
This surface instability was discovered by Higashitani\cite{Hig} in
connection with a study of the paramagnetic response of the MGS to
external magnetic fields. Our present calculation shows that the
surface instability is not sensitive to the spatial profile of the
gap; since we considered a finite value of the parameter
$\xi_0/\lambda$ we were able to calculate the phase transition
temperature.\cite{HonSignote} Within the framework of the
quasiclassical Green's function technique analogous results were
recently obtained by Barash {\it et al.}\cite{BKK_prb00}

For orientations of the surface different from $\alpha=\pi/4$, MGS
exists only for trajectories satisfying
$\mbox{sgn}(\Delta\bar\Delta)=-1$. The surface current is then
reduced, leading to a continuous reduction of the phase transition
temperature,
$k_B T_{TRSB} = (1/6)(\xi_0/\lambda)\Delta_0
\left(\cos^3\delta\alpha - \sin^3\delta\alpha\right)$,
with misorientation
$\delta\alpha=|\pi/4-\alpha|\in\left[0,\pi/4\right]$. For small
misorientations, $\delta\alpha\ll 1$, the reduction is quadratic in
$\delta\alpha$ and the instability is not dramatically sensitive to
the exact orientation of the surface. Note, however, that the
instability is quite sensitive to surface
roughness.\cite{Hig,BKK_prb00}

In conclusion, we have studied time-reversal symmetry breaking at
Josephson tunnel junctions of $d$-wave superconductors assuming that a
subdominant component of the order parameter is absent. For the
$D\gg\xi_0/\lambda$ junction case, at temperatures $k_BT\ll D\Delta_0$
for which MGS contribute to the Josephson current with a term
proportional to the first power in $D$, the TRSB is due to spontaneous
establishment of a phase difference $\phi=\pm\pi/2$ across the
junction which splits the MGS and produces a surface current. In the
extreme low transparency limit, $D\ll\xi_0/\lambda$, the mechanism
responsible for the instability is instead the same as for the free
surface. In this case, below a phase transition temperature
$k_BT_{TRSB}=(1/6)(\xi_0/\lambda)\Delta_0$, it is energetically
favorable to have Doppler shifted MGS carrying a surface current and
associated screening currents upholding the Doppler shifts. The
detailed spatial shape of the gap does not influence this instability.

\begin{acknowledgments}
  It is a pleasure to thank S. \"Ostlund for valuable discussions. We
  also thank Yu. S. Barash for pointing out Ref.
  \onlinecite{BKK_prb00} to us. Support from NFR and NUTEK, Sweden, is
  gratefully acknowledged.
\end{acknowledgments}

\newpage

\begin{thebibliography}{}
\bibitem{HonSignote}We note that in Ref.~\onlinecite{HonSig}, Honerkamp and Sigrist mentioned that the phase transition temperature should be of the order $(\xi_0/\lambda)T_{c}$.
\end{thebibliography}
\end{document}